# Observation of Bulk Polarization Transitions and Higher-Order Embedded Topological Eigenstates for Sound

Xiang Ni[1,2], Matthew Weiner[1,2], Andrea Alù[3,2,1], and Alexander B. Khanikaev[1,2]

[1]Department of Electrical Engineering, Grove School of Engineering, City College of the City University of New York, 140th Street and Convent Avenue, New York, NY 10031, USA.
[2]Physics Program, Graduate Center of the City University of New York, New York, NY 10016, USA.
[3]Photonics Initiative, Advanced Science Research Center, City University of New York, New York, NY 10031, USA

**Abstract:** Topological systems are inherently robust to disorder and continuous perturbations, resulting in dissipation-free edge transport of electrons in quantum solids, or reflectionless guiding of photons and phonons in classical wave systems characterized by Chern or spin-Chern topological invariants. These established examples of topological physics, however, do not exhaust all possible topological phases, and recently a new class of topological metamaterials characterized by bulk polarization has been introduced. In addition to edge conduction, these systems have been shown to host higher-order topological modes. Here, we introduce and measure topological bulk polarization in 3D printed two-dimensional acoustic meta-structures, and observe topological transitions as the design parameters are tuned. We also demonstrate that our topological meta-structure hosts both 1D edge and higher-order 0D corner states with unique acoustic properties. The edge states have spin polarization that reverses for opposite propagation direction, thus supporting directional excitation. Corner states are pinned to the meta-structure corners, and rapidly decay both along the edges and into the bulk. Interestingly, these 0D states can spectrally overlap with the continuum of bulk states, but are not compatible with radiation, thus enabling embedded topological eigenstates within the continuum of bulk modes. Their confinement and inherent topological robustness is experimentally confirmed by deliberately introducing disorder. Our findings open new directions in acoustics for advanced sound propagation and manipulation.

Topological phases of matter support some of the most fascinating properties for signal transport and wave propagation, holding the promise of revolutionizing technologies from quantum electronics (*1-7*) to photonics (*8-25*) and acoustics (*26-30*). In electronics and quantum photonics, they have been opening novel approaches for quantum computing interfaces (*31, 32*) and lasing (*33-35*), while in classical optics, mechanical and acoustic systems, they offer unprecedented robustness to defects and disorder (*10-20, 22, 24, 26-29, 36*). Most of the topological systems studied so far are characterized by integer topological invariants, such as Chern-class, Z2 invariants and winding numbers. More recently, a new class of symmetry-protected topological phases, characterized by bulk polarization, have been theoretically introduced (*37*). One example of such systems is given by quadrupole topological insulators (*38-41*), which have been recently implemented in mechanical (*42*) and microwave (*43*) systems, and in electrical circuits (*44*).

The class of topological systems with nontrivial bulk polarization offers a new opportunity to implement robust, precisely controllable physical responses. Additionally, as we show in the following, it enables the engineering and control of a wide variety of unique transport properties.

Specifically, we demonstrate that topological acoustic edge states with nontrivial bulk polarization support angular-momentum (pseudo-spin) to momentum locking, a property exclusively attributed to Chern class systems to date. These systems can be therefore used for precise directional excitation of the topological sound states using angular-momentum polarized sources. In addition, we show strongly localized higher-order corner states, despite being compatible to bulk propagation in terms of momentum, introducing the concept of topological embedded eigenstates that support high density of states and exhibit unique robustness to disorder. These findings open unprecedented opportunities for a new class of acoustic and photonic devices based on topological concepts.

**Topological acoustic metamaterial based on fractional bulk polarization**

We explore a two-dimensional acoustic Kagome lattice (*45, 46*) characterized by fractional bulk polarization of 1/3, supporting topological phase protected by $C_3$ symmetry. The lattice is schematically shown in Fig. 1A and it is formed by an array of acoustic resonator trimers coupled via narrow rectangular channels (Figs. 1B and 1C). Each resonator hosts acoustic pressure modes oscillating in the axial direction. We choose to work with the fundamental mode (~4.23 kHz), which has its only node at the center of the resonant cavity. The coupling strength is tuned by shifting the channels closer or farther away from the center node, thus enabling fine control over the local coupling strength. Due to the strong confinement of the resonant modes and the connectivity of the lattice, the system is well-approximated by the tight-binding model (TBM), with nearest-neighbor coupling described by inter-cell $\gamma$ and intra-cell $\kappa$ coupling parameters (Fig. 1A). For the case of an ideal Kagome lattice, with $\gamma = \kappa$, the band diagram obtained with TBM (grey solid lines in Fig. 1D) supports a Dirac-like degeneracy at the $K$ and $K'$ points. The degeneracy is formed between low-frequency monopolar mode, characterized by in-phase vibrations in all three cylinders of the trimer, and dipolar bands, which are left- and right-handed circularly polarized at the K and K' points, respectively.

As the symmetry is broken by detuning inter-cell $\gamma$ and intra-cell $\kappa$ couplings between neighboring trimers, a topological transition emerges in the band diagram. Specifically, the trimers in Figs. 1B and 1C (geometry provided in the Methods section) support bandgaps at both K and K' valleys with topologically nontrivial ($\gamma > \kappa$) and trivial ($\gamma < \kappa$) nature (referred to as expanded and shrunken respectively), implying that a control over the coupling parameters can enable ad-hoc topological transitions. The symmetry reduction from 6-fold rotational symmetry ($C_6$) to 3-fold rotational symmetry ($C_3$) leads to hybridization and avoided crossing of formerly degenerate dipolar and monopolar bands, giving rise to the band inversion. The cases of equal detuning for $\gamma > \kappa$ and $\gamma < \kappa$ have identical band structure, shown in Fig. 1D by solid lines, and therefore cannot be distinguished. However, these two cases represent two distinct topological phases, separated by the gapless $\gamma = \kappa$ transition point. The topological transition is demonstrated by directly calculating the bulk polarization (see supplementary Fig. S1) through a Wilson loop, and also by investigating the $C_3$ related properties at high symmetry points in the Brillouin zone, as we discuss in the following. We stress that the geometry of Fig. 1 can be viewed as the 2D counterpart of the well-known one-dimensional Su-Schrieffer-Heeger (SSH) model. In particular, just as in the SSH dimerized lattice, the topological or trivial character of the model is defined by the relative magnitude of the coupling coefficients in the unit cell. Additionally, the bulk polarization exhibits transition between trivial and nontrivial due to the change in choice of the unit cell (by a vector **r**, as shown in Fig. 1A). This ambiguity is removed only when a finite structure is considered, as a boundary is introduced. For a detailed explanation of the polarization difference in the trivial and topological scenario, the interested reader can refer to Supplementary Section S5.

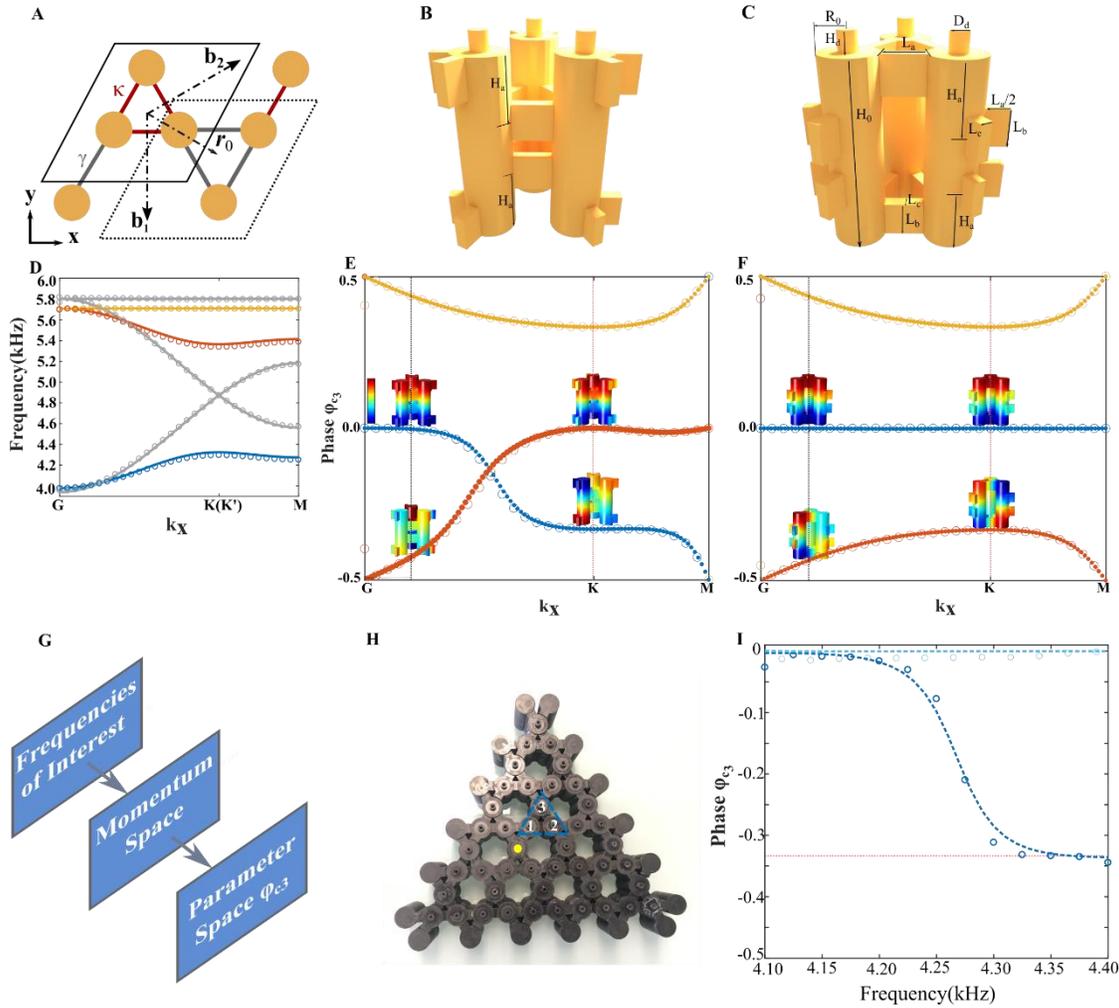

**Fig. 1. Concept and measurement of bulk polarization transition in the deformed Kagome lattice.** (**A**) Schematics of the 2D tight-binding Kagome lattice and (**B**-**C**) realistic acoustic trimer for "expanded" and "shrunken" lattice designs, which emulate the effect of nearest neighbor coupling. (**D**) Band structures of the Kagome lattice obtained by TBM (solid lines) and first-principle calculation (circular dots), separately. Normalized $\text{Arg}(\theta_n(k))$, $\phi_{C_3}$, along the high-symmetry point lines of (**E**) expanded lattice and (**F**) shrunken lattice. (**G**) Schematic explanation of one-to-one mapping from frequency space to parameter space of $\phi_{C_3}$. (**H**) Photograph of the printed acoustic lattice, the array consists of 5 cells at each edge with boundary cells that have a higher resonance frequency to terminate the array. A speaker is placed at the yellow spot, phase measurement are taken in the triangular region marked by blue lines. (**I**) $\phi_{C_3}$ of the lowest bulk band for topological nontrivial (dark blue) and trivial geometries (light blue). Both theoretical and experimental results are shown, denoted by dashed line and circle-shaped markers, respectively.

In addition to the results from TBM, the circular markers plotted in Fig. 1D illustrate first-principle simulations obtained by directly solving the three-dimensional pressure-wave equation using the finite element method (FEM). The accurate agreement between analytical and numerical results serves as further evidence of the applicability of TBM and the related theoretical analysis.

To confirm the topological transition as we tune the coupling, going from expanded to shrunken lattice, we extracted the topological polarization using both TBM and FEM. As shown in Supplementary Section S3, the polarization is directly related to the constraint of symmetry operations on the eigenvectors. In particular, when $C_3$ symmetry is preserved we can extract the polarization directly from the eigenvalue solution, as detailed in Supplementary Section S4. The bulk polarization assumes the form

$$e^{-i\pi(p_i)} = \prod_{n \in occ} \frac{\theta_n(K)}{\theta_n(\Gamma)}, \qquad (1)$$

where $\theta_n(\mathbf{k}) = \langle u_n(\mathbf{k})|R_3|u_n(\mathbf{k})\rangle$ is the expectation value of the $C_3$ group operator $R_3$ (rotation by $2\pi/3$) applied to eigenvector $u_n(\mathbf{k})$. The index $i$ corresponds to each reciprocal vector in the Brillouin zone and the phase of $\theta_n(\mathbf{k})$, function of momentum vector $\mathbf{k}$, namely $\phi_{C_3}(\mathbf{k})$, is plotted in Fig. 1E and F, clearly revealing the difference between $\gamma > \kappa$ and $\gamma < \kappa$. This approach allows a straightforward physical interpretation of topological phase: in the topological nontrivial regime, the $R_3$ operator acting on the lower band near the K (or K') point yields the eigenvalue $2\pi/3$, associated with the rotational dipolar field-profile (blue line in Fig. 1E). In the topologically trivial regime, the pattern is not rotating in the entire Brillouin zone, giving rise to no phase difference between K and $\Gamma$ points (blue line in Fig. 1F).

Interestingly, this approach enables an alternative way of accurately extracting the bulk polarization by only evaluating $\theta_n(\mathbf{k})$ at $C_3$-invariant points of the Brillouin zone, because in this case they are also eigenvalues of $R_3$. The bulk polarization of the band, which characterizes the topological phase of interest, can therefore be defined as the difference of $\phi_{C_3}(\mathbf{k})$ at different $C_3$-invariant points, K and $\Gamma$ points respectively, i.e., $p_i = \phi_{C_3}(K) - \phi_{C_3}(\Gamma)$ for the lowest band of interest. For the lower-frequency (blue-colored) band, the bulk polarizations can be readily seen from Figs. 1E and 1F as

$$(p_1, p_2) = \begin{cases} \left(-\frac{1}{3}, -\frac{1}{3}\right), \kappa < \gamma, \\ (0,0), \kappa > \gamma, \end{cases} \qquad (2)$$

which clearly indicate distinct topologically nontrivial and trivial phases, respectively.

In order to extract experimentally the bulk polarization in a finite acoustic meta-structure, we exploit the bijective relations between frequency, momentum space, and the $C_3$ symmetric portrait of the bulk mode, which is valid for an isolated low-frequency band, as schematically depicted in Fig. 1G. These relations hold true due to the fact that the corresponding bulk mode has the lowest frequency at $\Gamma$-point, and the highest frequency at K(K')-point. In addition, the relative direction connecting source (speaker) and detector (microphone) can be selected so that waves with either positive or negative momentum vectors are probed. Therefore, due to the one-one correspondence between the desired frequencies and high symmetric points, the mode at Γ-point can be probed by a low-frequency excitation, whereas the modes at the K(K′)-point can be probed by driving the system with high-frequencies (close to the lower-frequency edge of the topological bandgap).

The acoustic Kagome lattices of Fig. 1 were fabricated using a B9Creations stereolithographic 3D printer (see Methods). As shown in Fig. 1H, the source was placed at the center of the structure (indicated by the yellow dot), and two detectors were placed inside blue trimers of either topological nontrivial or trivial lattices. The relative direction between source and detectors corresponds to the case of excitation with either positive or negative momentum. Thus, by sweeping the frequency of the source, we can measure the field profile of the excited bulk mode, obtain the phase within the trimers, reconstruct the eigenstate corresponding to this mode, and extract the phase $\phi_{C_3}$. Details of phase measurement and data process are described in the method section. As shown in Fig. 1I, the measurement results (circular markers) and first-principle simulations (dashed lines) clearly show that the phase $\phi_{C_3}$ sweeps 1/3 of $2\pi$ for the case of the topological lattice and remains at 0 for the topologically trivial lattice. These results unambiguously confirm and experimentally verify the fractional topological polarization of 1/3 of the corresponding low-frequency bulk band for the expanded lattice (red trimer in Fig. 1H).

**Edge states and Corner states**

The most prominent distinction between topological and trivial phases of matter is in the robust emergence of topological edge states exponentially confined to the boundaries. As seen from Fig. 2A, blue region, the triangular Kagome lattice in the topological regime supports topological edge states, which emerge from the non-vanishing bulk polarization (Fig. 1E), contrary to the edge states due to nontrivial Berry curvature in QSHE and VHE systems. In addition to edge states, the energy spectrum (red line in Fig. 2A) reveals another class of totally dispersionless states, which suggests their 0-dimentional nature. These states indeed represent a new class of higher-order (D-2) topological states, which are confined to corners of the system when the angle of the corner equals 60 degrees. For all values of parameter $\kappa$ (and fixed γ), these states are pinned to "zero-energy" which, for this acoustic system, represents the frequency of the isolated acoustic resonator.

Interestingly, the corner states also exhibit an intriguing distinction from their cousins found in quadrupolar topological insulators (*38*). Apart from their topological charge of 1/3 (as opposed to 1/4 in quadrupolar insulators), these states exist in two distinct regimes. In the first regime, when the symmetry reduction is large, and $\kappa/\gamma < 0.5$, these states are embedded in the band gap induced by the topological transition (white region on the right side of Fig. 2A). In the second regime, when $1 > \kappa/\gamma > 0.5$, symmetry breaking is not sufficient to lift a bulk gap wide enough and, as a result, the corner states (one state per corner) are embedded into the bulk continuum, becoming embedded eigenstates (*47-51*). Our numerical calculations of the supercell demonstrate that these 0D topological eigenstates do not interact with bulk states, despite being compatible to radiation in the bulk in terms of momentum, and multiple degeneracies (crossings) between bulk and corner spectra are observed. In all regimes, the corner states exhibit exponential decay with plane-wave-like modulation in the direction bisecting the corner.

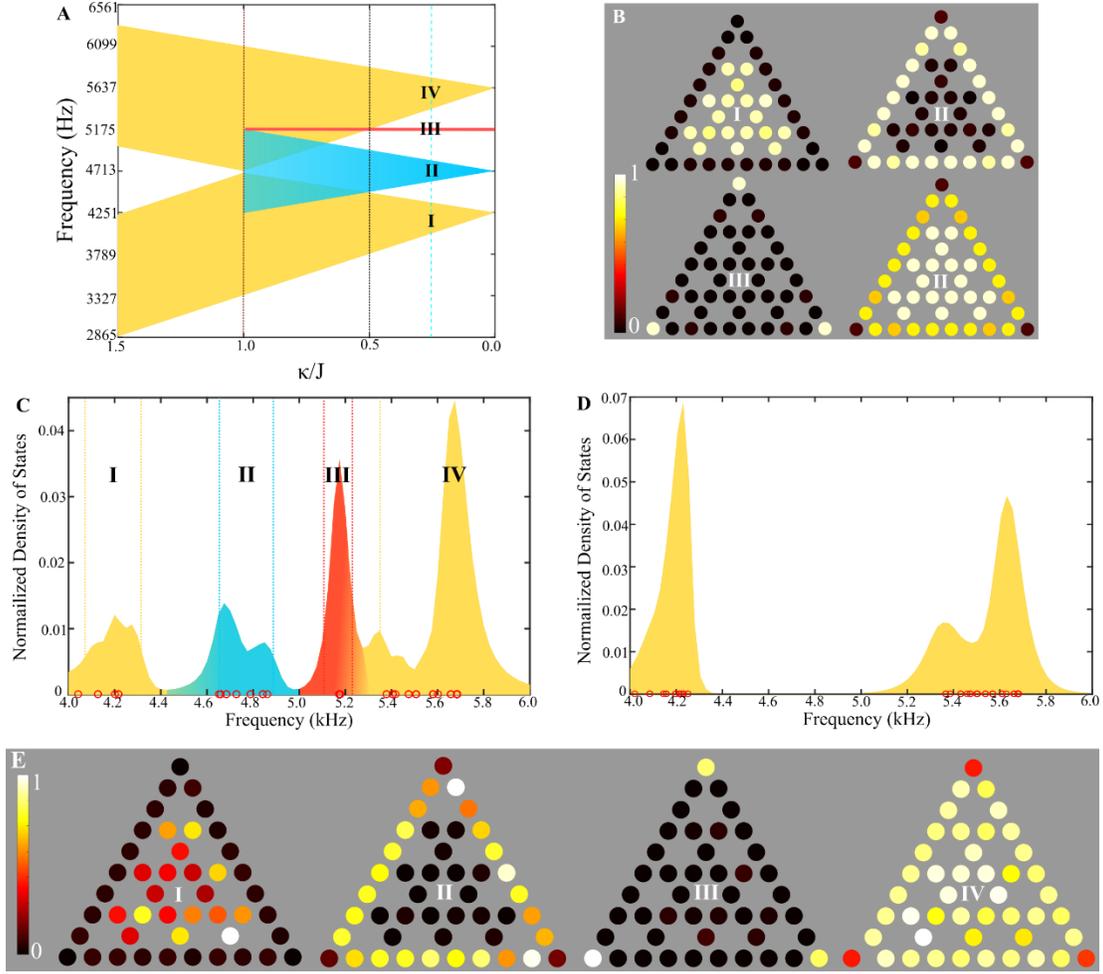

**Fig.2 Theoretical prediction and experimental demonstration of acoustic edge and corner states.** (**A**) Energy spectrum calculated from TBM for the triangular Kagome lattice. Yellow, blue and red bands are for bulk, edge and corner states, respectively. (**B**) Normalized spatial distribution of $|\psi(\omega)|^2$ integrated over the frequencies of lower bulk, edge, corner and upper bulk regions, respectively, when $\kappa/\gamma = 0.252$ as denoted by blue dashed line. $\psi(\omega)$ is the eigenstate of the triangular Kagome lattice. (**C**) Normalized density of states for the expanded lattice obtained from the measurements of acoustic power at the top of the trimer. Red, blue and yellow regions are color-coded to represent corner, edge and bulk modes dominating in these regions. Red, blue and yellow dashed vertical lines show lower and upper bounds of integrations used in **E**. (**D**) Normalized density of states for shrunken lattice obtained from the acoustic power measured at the top of the trimer. Results from numerical calculation are marked in the red circle in both (**C**) and (**D**). (**E**) Spatial distributions of acoustic power integrated over the respective frequency regions indexed by roman numerals (from I to IV).

When the ratio $\kappa/\gamma$ is equal to 0.252, as shown by the blue dashed line in Fig. 2A, bulk, edge and corner modes are well separated from each other. To verify the localization of edge states and corner states, the spatially resolved eigenstate distributions $|\psi(\omega)|^2$ integrated over the corresponding frequencies (indexed by I-IV) are plotted in Fig. 2B. To reveal all the states hosted by both the topological and trivial structures, we measured the local density of states, as shown in Figs. 2C and 2D, respectively. The density of states was extracted by studying the local response through exciting and probing pressure field at the same location at each lattice site. The measurement was performed over a broad frequency range that enclosed both low- and high-frequency bulk bands. In agreement with the results in Fig. 2B, the measured field profiles corresponding to the excitation of the low-frequency bulk, edge, corner and high-frequency bulk states (Fig. 2E) clearly highlight the different topological nature of the different bands. As expected, the spatially resolved power distribution integrated over the respective frequency ranges is directly related with the peaks in the measured density of states distribution in Fig. 2C, corresponding to the excitation of edge (blue shaded region) and corner (red shaded region) states, which arise only in the topological lattice. As seen from the field profiles of the corner states, where only one of the three sites of the trimer is excited, the state carries a topological charge of 1/3.

For the given set of measurements, the considered detuning of the off-site and on-site coupling was strong, and both corner and edge states therefore appeared in the topological bandgap. The observed small overlap of the corner states with upper bulk modes is explained as the result of inhomogeneous broadening caused by the finite lifetime of the modes in the experimental setup. We found that the main mechanisms contributing to the broadening were the geometrical deviations the due to limited fabrication precision, as well as radiative-loss through the probe (excitation) channels. Nonetheless, as seen from the comparison with the theoretical data shown in circles on the horizontal axis of Figs. 2C and 2D, the experimental data are in very good agreement with first-principle results, indicating that both factors do not affect the topological nature of the modes, confirming their robustness, and do not alter their spectral position.

**Angular momentum (pseudo-spin) to momentum locking of topological edge states**

Topological edge states have been shown to have remarkable properties associated with their nature. In particular, the robustness of topological states in TR-symmetric systems has been attributed to spin-momentum locking (QSHE) (*22, 36, 52, 53*). Despite the fact that the system presented here belongs to a completely different topological class, we found a similar locking associated to angular momentum (pseudo-spin) in the edge states of our system. This property stems from the rich structure of bulk modes mentioned above. Thus, due to the circularly polarized character of the bulk bands, which are circular right-handed near K point and circular left-handed near K' point, the edge states also exhibit pseudo-spin polarization which reverses for opposite wavenumbers. To corroborate this argument, we conducted a first principle calculation of the supercell in Fig. 3A. The calculated band structure reveals two types of edge states, localized at the lower end (blue band) and upper end (red band) of the strip (Fig. 3B). A careful analytical study of the topological edge states is presented in Supplementary Section S8.

The phase $\phi_{C_3}$ of the edge states extracted from the eigenstate distributions over the corresponding edge trimers is shown in Fig. 3C. Though the red edge band barely exhibits a circular polarized behavior, the blue edge band does behave like a circularly polarized mode when the projected momentum vector $k_x$ approaches the boundary of the Brillouin zone ($k_x = \pm\pi$). The 3D printed triangular-shaped lattice has the same cut at three boundaries, and therefore it only holds the edge states of interest, lying in the blue band. Again, using the principle of one-one correspondence among frequency, momentum and phase $\phi_{C_3}$, we performed a frequency sweep over the range of blue edge band (shown as the light green shaded region in Fig. 3B) with sources carrying different angular momentum. Indeed, due to pseudo-spin locking, the edge modes are excited unidirectionally, traveling in different directions as a function of the source handedness. As experimentally demonstrated below, this feature enables robust directional excitation of topological edge states.

We next experimentally investigated the nature of these edge states. To this aim we measured the transmission along the edge by placing the source at the corner of both topological and trivial structures, and measured the signal by placing a detector at the center of the opposite edge. The measurement results for both trivial and nontrivial structures are shown in Fig. 3D, clearly revealing the presence of a bandgap in both cases and an additional midgap transmission peak associated with the excitation of edge states in the topological lattice. The measured frequency range of edge states is consistent with our theoretical predictions (shaded green region).

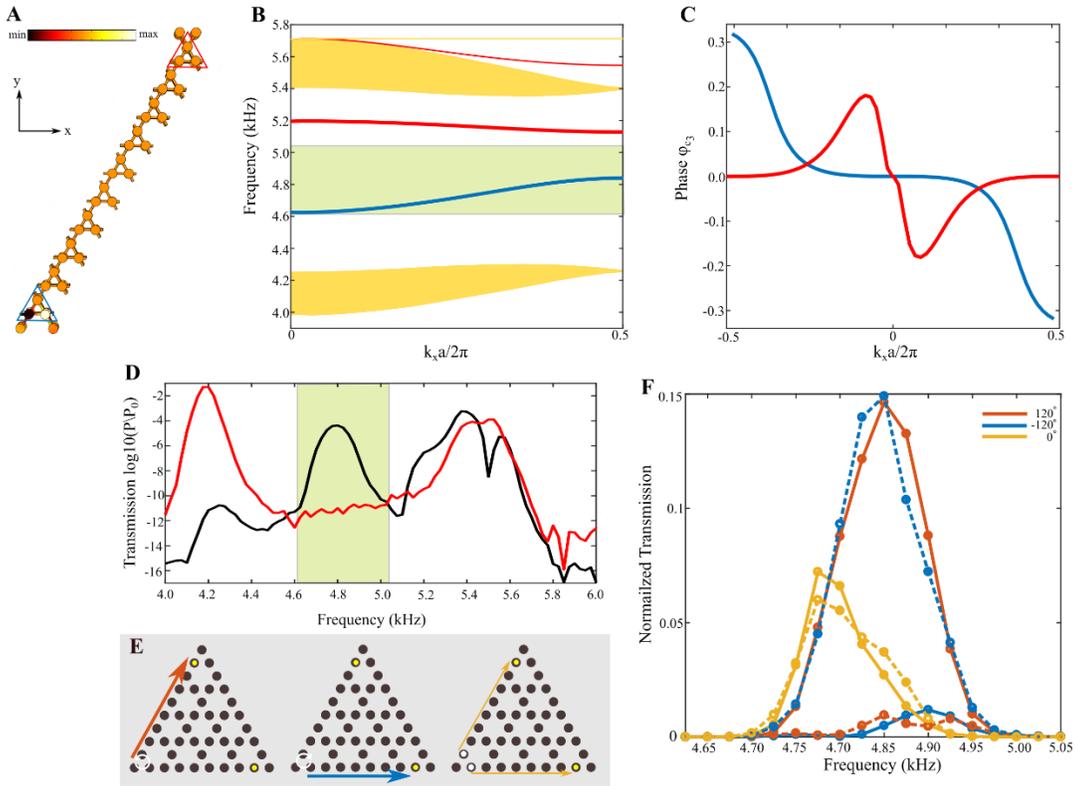

**Fig. 3. Experimental demonstration of angular momentum (pseudo-spin) locking of the edge states.** (**A**) Field distribution of lower edge band in the supercell consisting of 10 cells near $k_x = \pi$, the supercell is periodic in the *x* direction and terminated by a boundary cell in the *y* direction. (**B**) Band structures of a supercell terminated by cylinders with half height of the trimers. Red and blue bands denote edge states located at the upper and lower end of the supercell, and yellow bands are for bulk states. (**C**) Normalized $\phi_{C_3}$ of the lower (blue) and upper (red) edge band, calculated for the edge states of the supercell, denoted by blue and red triangle in (**A**), respectively. (**D**) Transmission spectra for the case of expanded lattice (black line) and shrunken lattice (red line), respectively. The shaded light green region indicates the frequency region of the edge state of interest. (**E**) Schematics of edge state excitation by different sources;

the sources from left to right are left-handed circular polarized, right-handed circular polarized and linearly polarized. Yellow dots are the probing positions and white dots are the source position. (**F**) Normalized transmission spectra measured at two edges illustrated in (**E**), for phase differences between the two sources of 120°, −120° and 0°, respectively. Solid and dashed lines are the spectra measured at the upper left and bottom right yellow dots in (**E**).

To further explore the properties of these acoustic edge modes and to exploit their angular momentum to momentum locking, we placed two sources with controllable phase shift at the two sites of the same corner trimer, as shown in Fig. 3E, and measured the amplitude response as the frequency was swept over the range of the blue edge band at the upper left edge and the bottom right edge of the structure (yellow dots in Fig. 3E). We found that the transmissions for the two sites became strongly dependent on the relative phase in the frequency range of edge states. When the phase difference of the sources is 120°, which mimics a left-handed circularly polarized source, stronger excitation occurs at the upper left edge compared to the excitation at the bottom right edge, as shown by red curves in Fig. 3F. Meanwhile, the maximum excitation is observed near the large frequency end of the edge band (~4.85kHz), implying that the edge mode with left-handed circular polarization near the projected M-point is excited. Repeating the same measurement when the phase difference is −120°, mimicking a right-handed circularly polarized source, stronger excitation occurs at the bottom right site (blue curves in Fig. 3F), thus confirming the rotating character of the modes and the locking of angular momentum to the propagation direction. Interestingly, when the phase difference is 0°, transmissions at two sites are detected with nearly the same magnitude, with a lower maximum value compared to the ones of circular polarized sources, indicating that the energy flux equally splits towards two edges, because of symmetry. In this case, the transmission is maximum at the lower frequency, and it is suppressed at the higher frequency because of the non-circular polarization of the source, which is consistent with our theoretical prediction in Fig. 3C.

**Robustness of the corner states**

The nontrivial bulk polarization of the Kagome lattice suggests that the corner states may possess some degree of protection against disorder. To confirm this, we selected again the case when $\kappa/\gamma = 0.252$ and performed TBM studies of disordered systems with all the cylinders, except for the corner one, having the resonant frequency randomly distributed around the frequency of the corner cylinder. We found that the corner state remains pinned to the same frequency for small to moderate levels of disorder, an evident effect of strong confinement and robustness. Even for levels of disorder as large as 16%, the corner states are well isolated from the bulk and edge states (Fig 4A). On the other hand, the frequency of the corner state was very sensitive to the local frequency of the resonator it was pinned to, thus suggesting possible applications for sensing.

We also experimentally investigated the properties of the corner states. As suggested by theoretical results, these modes should exhibit robustness against disorder. To test this, we fabricated several trimers (green dashed line region in Fig. 4B) with deliberately introduced random deviation of up to 10% in their height, and placed them next to the corner trimer, in which two sites were randomly perturbed to the same degree, whereas the exact corner site remained unaltered. As a result, the resonant frequency of the individual cylinders fluctuated randomly by up to 400Hz, which is about 50% of the spectral width of the topological bandgap. The spatially resolved measurements of the pressure field shown in Fig. 4B confirm that the corner states retain a strong confinement, and do not leak into either bulk or edge states.

In a second measurement, the three corner trimers of the printed structure were perturbed by a slight deviation of up to 10% in their height, except the exact corner sites which were left unperturbed. Corner states were still well localized at their corners, as shown by the spatially resolved power distribution integrated over the frequency range of the corner modes in Fig. 4C. The measured density of states confirms that the corner states remain well defined in the spectrum, and overlap with the corner states for the unperturbed lattice, indicating their robustness against strong disorder and imperfections (Fig. 4D).

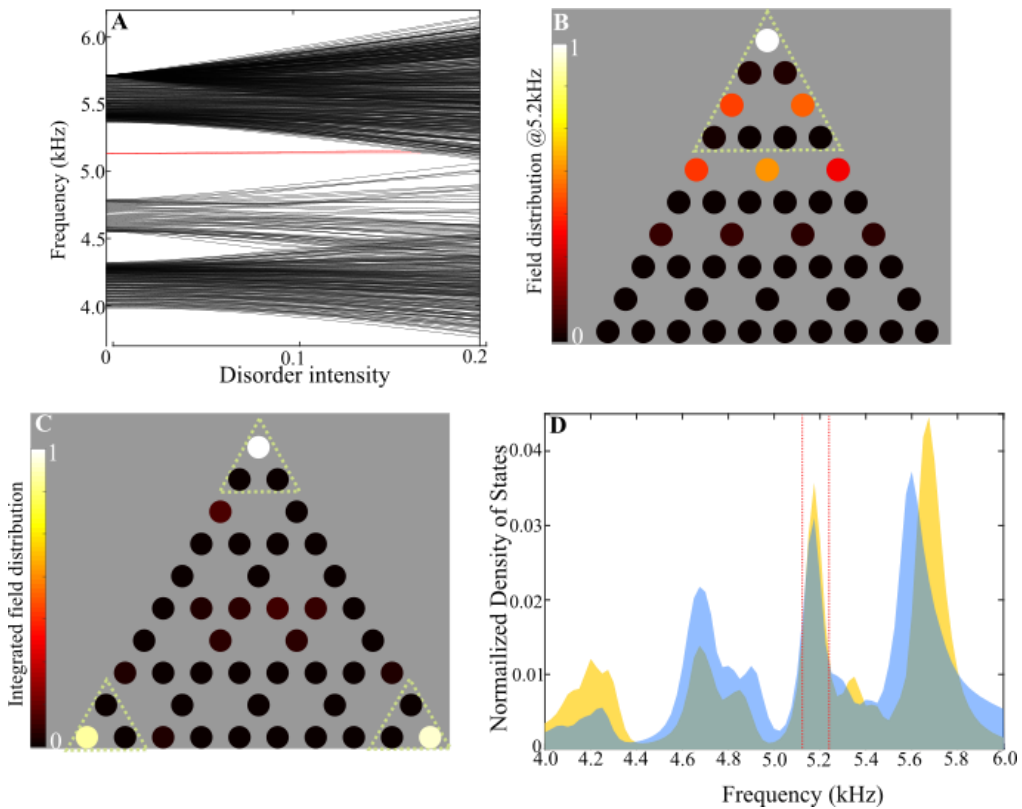

Fig. 4 **Effect of disorder and robustness of the corner states.** A. TBM results demonstrating the pinned character of the corner state in the presence of disorder. Triangular Kagome lattice consisting of 20 trimers at each site, all sites except the three corner sites are perturbed by small deviation in their height, and these disorder parameters are randomly generated. The disorder intensity represents the ratio of the deviation over the unperturbed height. B. Spatially resolved field distribution when the source was put at the corner with frequency around the onsite frequency of the corner states. C. Field distribution integrated over the frequency range of corner states, indicated by the red dash lines in D. The disorder was introduced at the three corner trimers, marked by light green dashed lines. D. Normalized density of states for the disordered lattice in C,

marked by the blue shaded curve, and density of states for non-disordered one also plotted in yellow shaded curve.

**Topological corner states embedded in the continuum**

Next, we studied the case of corner states embedded in the continuum of bulk states. To this aim, we fabricated another topological sample with detuning between inter-cell and intra-cell channels such that the corner states would spectrally overlap with high-frequency bulk modes, corresponding to the ratio $\kappa/\gamma = 0.52$ in Fig. 2A. The corner states were probed by placing a source at the three corner sites of the topological structure, which were driven at 5250Hz. The bulk states were excited by placing the source inside the structure. Results shown in Fig. 5A-C prove that the corner states remain localized to the corners and do not radiate into the bulk. At the same time, the source, at the same frequency, excited predominantly bulk states when placed inside the structure. Only one corner state was excited weakly in this case due to some spatial overlap of the pressure field produced by the source with the one of the exponentially localized corner states. These results prove that, despite their spectral overlap and momentum compatibility, the discrete corner states remain embedded into the bulk continuum without coupling into it, opening a door for studying ultrasharp Fano resonances with topological discrete states, when coupling with the continuum is controllably introduced by deliberate symmetry reduction.

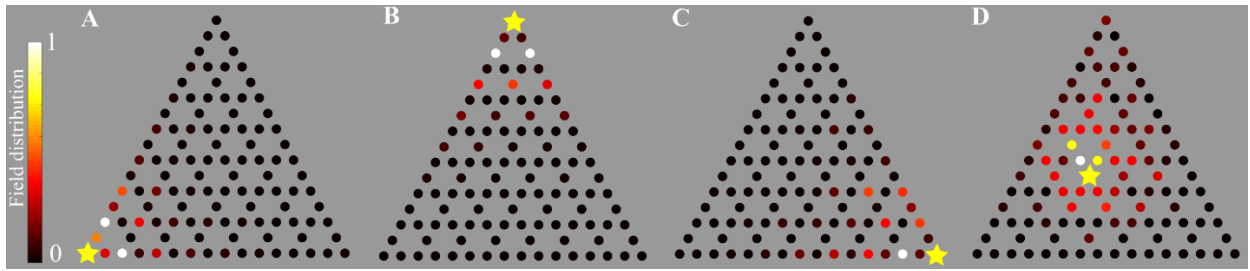

**Fig. 5 Experimental demonstration of corner states coexisting with continuum of bulk modes.** Field distribution in the lattice caused by the source located (as indicated by the yellow star) at (**A**) bottom left corner, (**B**) top corner, (**C**) bottom right corner, or (**D**) in the bulk. The excitation frequency is ~5250Hz. Due to narrower topological bandgap and weaker localization of the corner states, the size of the fabricated structure was increased to 8 trimers for each edge.

**Discussion**

The new class of topological states introduced in this paper has a great potential for applications over a broad range of disciplines, from photonics to acoustics and mechanical vibrations. Indeed, our work shows that some of the topological features attributed solely to Chern class topological systems (such as spin-momentum locking) can be induced in Kagome lattices with synthetic degrees of freedom. In our case, the control over angular momentum allows for precise steering of radiation. Thus, synthetic degrees of freedom bring an entirely new dimension to fields lacking internal degrees of freedom, such as acoustics. We have also shown that these synthetic degrees of freedom can be beneficial to generate nontrivial bulk polarization states, for the design of systems supporting unconventional edge and corner states. Unique features, such as angular-momentum to momentum locking, robustness, and the non-radiative character of higher-order topological embedded eigenstates significantly expand our ability to control, trap, and steer waves, enabling a plethora of novel applications, from robust tunable waveguides and beam splitters to a new generation of high fidelity topological sensors. We believe that our work opens important directions in acoustics and beyond, offering experimental evidence for new ways to control sound in unprecedented ways.

**Methods**

*Structure design, 3D printing, and generic measurements* – The unit cell designs of the topological lattice and trivial lattice are plotted in Fig. 1B-C, the lattice constant of the structure is $a_0 = 39.75mm$, and the height of cylinder is chosen as $H_0 = 40.52mm$, with the radius $R_0 = 5.77$mm, such that the frequencies of the desired mode are in the probing range of the microphone. The connectors between the cylinders consist of blocks, with their dimensional sizes $L_a = 8.34mm, L_b = 7.95mm$ and $L_c = 2.64mm$, respectively. The coupling strength of the modes is maximal when the connectors are at the top or bottom of the cylinders, and minimum at the center of the cylinders. In order to make the intra-cell and inter-cell coupling of the trimers inequivalent, the outer connectors of the topological trimer shown in Fig. 1B are placed at the top and bottom of the cylinders, while the inner connectors are shifted toward the center by the distance $H_a = 10.23mm$. The trivial trimer, in Fig. 1C, has the opposite way around. The boundary cells, shown in the photograph of the fabricated structure (Fig. 1H), have the same parameters as the trimers, except their heights are half the trimers' height.

The trimers and boundary cells were fabricated using the B9Creator v1.2 3D printer. All cells were made with acrylic-based light-activated resin, a type of plastic that hardens when exposed to UV light. Each cell was printed with a sufficient thickness to ensure a hard wall boundary condition and narrow probe channels were intentionally introduced on top and bottom sides of each of the cylinders to excite and measure local pressure field at each site. The diameter of the port is $D_0 = 3.73$ mm, and the upper port has a height $H_d = 3.97$mm, while the height of lower port is $2H_d$. When not in use, the probe channels were closed with 3D printed cups. Each trimer and boundary cells were printed one at a time and the models were designed specifically to interlock tightly with each other. The non-trivial and trivial structures shown in Fig. 1 both consist of 15 trimers and 18 boundary cells. For the lattice studying corner states in the continuum in Fig. 5, 36 trimers and 27 boundary cells are used. For all measurements, a frequency generator and FFT spectrum analyzer scripted in LabVIEW were used. The FFT spectrum analyzer is also capable of extracting phase differences between two channels.

*Numerical method* –TBM is used to fit the band diagram of the topological and trivial structures and also to verify the correctness of $\phi_{C_3}$ graph based on the eigenstates calculation. For the gapless band diagram, the onsite frequency is fitted as $\omega_0 = 5182.5$Hz, and the coupling strength as $\kappa = \gamma = -310$Hz. For the topological band diagram, these fitted parameters are $\omega_0 = 5133$Hz, $\kappa = 115.5$Hz and $\gamma = 462$Hz and for trivial band diagram, onsite frequency is the same as the topological band diagram with the magnitudes of $\kappa$ and $\gamma$ flipped.

*Phase difference measurement* -  A set of two compact magnetic transducers was used to excite local pressure fields and two directional microphones (Model:EMM-6) were connected to an external two-port digital data acquisition device (AUBIO BOX USB 96), enabling excitation of the desirable phase profiles and measurements of the phase across the structure. For the phase differences, all the upper ports were open to allow the modes to propagate with vast loss, thus minimizing the wave reflection at the boundary. The speaker was placed at the site in the center of the structure, and two microphones were placed at the two sites within a bulk trimer near to the speaker, as shown in Fig. 1 H. The frequency generator was used to run a sweep of monochromatic frequencies from 4000 Hz to 4550 Hz in 25 Hz intervals and with the dwell time of 4 seconds (to allow for stability of the profile) and determine the phase difference between the two channels at each frequency. One of the microphones was also switched to the third site in the same trimer to measure the second phase difference. Since the detected modes near the source behave like bulk modes of the infinite structure due to large loss (when all ports are open and leak), it is permitted to construct the normalized eigenvectors of the bulk modes by these measured phase differences. With the information of eigenvectors, phase $\phi_{C_3}$ is extracted, as shown by Fig. 1I.

*Density of states (DOS) measurement* - The speaker was placed at the bottom port and the microphone at the top port of the same site. A tiny gap was left between the speaker and the port to allow for the presence of reflection channels while the microphone was closely touched with the port to achieve the maximum absorption. The frequency generator was used to run a sweep from 3600 Hz to 6000 Hz in 25 Hz intervals and with the dwell time of 4 seconds while the FFT spectrum analyzer obtained the amplitude responses $\varphi(\omega)$ at each frequency. Field distributions $\varphi(i, \omega)$ are obtained by repeating this process for each site $i$. Since it's hard to guarantee the tiny gap is exactly the same for every site, and the amplitude response is highly sensitive to this tiny gap, we normalized the data for each site based on the total spectra summed over frequencies as well as on the free space amplitude response between the microphone and the speaker, $\Phi(i, \omega) = \varphi(i, \omega) / \sum_\omega \varphi(i, \omega) / \varphi_{\text{air}}(\omega)$. After that, we squared the signal $\Phi_n(i, \omega)$ and averaged the power spectrum for an array of $N$ resonators $P_a(\omega) = \sum_i |\Phi(i, \omega)|^2 / N$ to get the normalized spectra, $P_n(\omega) = P_a(\omega) / \sum_\omega P_a(\omega)$, the equivalent to the density of states of the Kagome lattice. For the field profiles excited by a single frequency, the speaker was fixed at the port of the site of interest and the microphone was placed over each site of the lattice to measure the magnitude response at the desired frequency (5250 Hz for the corner state).

*The disorder parameters* – In the first disorder configuration shown in Fig. 4A, the disorder parameters are generated by the random function in MATLAB script, and converted to the small deviation of the height $H_0$, the ratios of the perturbed height over $H_0$ were 1.00, 0.9076, 1.05, 0.9607, 0.9820, 1.0340, 0.9961, 0.9220, 1.0182. In the second disorder configuration in Fig. 4B, the disorder parameters are 1.00, 0.9076, 1.05; 1.00, 0.9744, 1.0273; 1.00, 0.9348, 1.043.

## Data availability

Data that are not already included in the paper and/or in the Supplementary Information are available on request from the authors.

## Author contributions

All authors contributed extensively to the work presented in this paper.

## Acknowledgements

The work was supported by the National Science Foundation grants CMMI-1537294 and EFRI-

1641069. Research carried out in part at the Center for Functional Nanomaterials, Brookhaven National Laboratory, which is supported by the U.S. Department of Energy, Office of Basic Energy Sciences, under Contract No. DE-SC0012704.

**Competing interests**
The authors declare no competing interests.

**Corresponding authors**
Correspondence to Andrea Alù or Alexander B. Khanikaev.

# Supplementary Information:
# Observation of Bulk Polarization Transitions and
# Higher-Order Embedded Topological Eigenstates for Sound

Xiang Ni[1,2], Matthew Weiner[1,2], Andrea Alù[3,2,1], and Alexander B. Khanikaev[1,2]

[1]Department of Electrical Engineering, Grove School of Engineering, City College of the City University of New York, 140th Street and Convent Avenue, New York, NY 10031, USA.
[2]Physics Program, Graduate Center of the City University of New York, New York, NY 10016, USA.
[3]Photonics Initiative, Advanced Science Research Center, City University of New York, New York, NY 10031, USA

## S1 Eigenvalues and eigenstates of Kagome lattice from tight binding model (TBM)

We start with a two-dimensional infinite Kagome lattice, in which the unit cell is defined by the rhombus region in Fig. 1A, with a lattice constant $a_0$. Here TBM is applied to the lattice and only the nearest hopping term is considered into the model and assumed as $t_1 = 1$. Consequently, the kernel of Hamiltonian in the momentum space is

$$\mathcal{H}(\boldsymbol{k}) = \begin{pmatrix} 0 & d_1 + id_2 & d_3 + id_4 \\ d_1 - id_2 & 0 & d_5 + id_6 \\ d_3 - id_4 & d_5 - id_6 & 0 \end{pmatrix}, \tag{S1}$$

where $d_1(\boldsymbol{k}) = -\kappa - \gamma \cos\left(\frac{(k_x+\sqrt{3}k_y)a_0}{2}\right)$, $d_2(\boldsymbol{k}) = -\gamma \sin\left(\frac{(k_x+\sqrt{3}k_y)a_0}{2}\right)$, $d_3(\boldsymbol{k}) = -\kappa - \gamma \cos\left(\frac{(-k_x+\sqrt{3}k_y)a_0}{2}\right)$, $d_4(\boldsymbol{k}) = -\gamma \sin\left(\frac{(-k_x+\sqrt{3}k_y)a_0}{2}\right)$, $d_5(\boldsymbol{k}) = -\kappa - \gamma \cos(k_x a_0)$, $d_6(\boldsymbol{k}) = -\gamma \sin(k_x a_0)$. The Eigenvalue equation is

$$\mathcal{H}(\boldsymbol{k})|u_{nm}(\boldsymbol{k})\rangle = \epsilon_n(\boldsymbol{k})|u_n(\boldsymbol{k})\rangle, \tag{S2}$$

in which $|u_n(\boldsymbol{k})\rangle = a_{n,1}(\boldsymbol{k})|1\rangle + a_{n,2}(\boldsymbol{k})|2\rangle + a_{n,3}(\boldsymbol{k})|3\rangle$ is the periodic function of the Bloch state, and $\kappa$ and $\gamma$ are intra-cell and inter-cell hopping terms, respectively. The eigenvalues and eigenvectors are obtained from (S2) are:

$$\epsilon_1(\boldsymbol{k}) = \frac{1}{2}\left(-\kappa - \gamma - \sqrt{9(\kappa-\gamma)^2 + 8\kappa\gamma\left(\cos(k_x) + 2\cos\left(\frac{k_x}{2}\right)\cos\left(\frac{\sqrt{3}k_y}{2}\right) + \frac{3}{2}\right)}\right),$$

$$\epsilon_2(\boldsymbol{k}) = \frac{1}{2}\left(-\kappa - \gamma + \sqrt{9(\kappa-\gamma)^2 + 8\kappa\gamma\left(\cos(k_x) + 2\cos\left(\frac{k_x}{2}\right)\cos\left(\frac{\sqrt{3}k_y}{2}\right) + \frac{3}{2}\right)}\right),$$

$$\epsilon_3 = -\kappa - \gamma, \tag{S3}$$

$$u_n(\mathbf{k}) = \begin{pmatrix} \epsilon_n(d_3 + id_4) + (d_1 + id_2)(d_5 + id_6) \\ \epsilon_n(d_5 + id_6) + (d_1 - id_2)(d_3 + id_4) \\ \epsilon_n^2 - (d_1^2 + d_2^2) \end{pmatrix}.$$

(S4)

With the eigenvalues and eigenstates of system in hand, we can easily study the $C_3$ symmetric related properties at high symmetric points and bulk polarization as explained later.

**S2 Wannier center and bulk polarization of Kagome lattice**

It is well known the eigenvalue problem of the Wilson loops can be expressed as (1-3)

$$W_{k_s+2\pi \leftarrow k_s, k_t} |v_{\mathbf{k}}\rangle^j = e^{i2\pi v_s^j(k_t)} |v_{\mathbf{k}}\rangle^j,$$

(S5)

where Wilson loop is defined as $W_{2\pi+k_s \leftarrow k_s, k_t} = \langle u_{2\pi+k_s,k_t} | u_{2\pi+k_s-\delta k,k_t} \rangle \langle u_{2\pi+k_s-\delta k,k_t} | u_{2\pi+k_s-2\delta k,k_t} \rangle \ldots \langle u_{k_s+2\delta k,k_t} | u_{k_s+\delta k,k_t} \rangle \langle u_{k_s+\delta k,k_t} | u_{k_s,k_t} \rangle$, and $k_s, k_t = 0, \delta k, \ldots, (N_k - 1)\delta k$, $\delta k = \frac{1}{N_k}\frac{4\pi}{\sqrt{3}a}$. $j$ represents the index of the occupied bands, $j = 1, 2, \ldots, N_o$. $v_s^j(k_t)$ is called the Wannier center of the lattice, and the polarization of the occupied bands can be obtained through the Wannier center,

$$p_s = 1/N_k \sum_{j, k_t} v_s^j(k_t).$$

(S6)

Following the above formula, the Wannier bands corresponding to the lowest energy band of expanded, unperturbed and shrunken Kagome lattices are plotted in Fig. S1 (A-C). Wannier bands for the directions of $\mathbf{b_1}, \mathbf{b_2}$ and $\mathbf{b_3}$ shown in Fig. S2. are exactly overlapped. Note for the unperturbed case, the Wannier band is discontinuous at the high symmetric points K(K') because of the band degeneracies. The bulk polarization of the expanded lattice calculated by (S6) is $-\frac{1}{3}$, thus, the topology is nontrivial. The bulk polarization of shrunken lattice, meanwhile, is 0, and it's topological trivial.

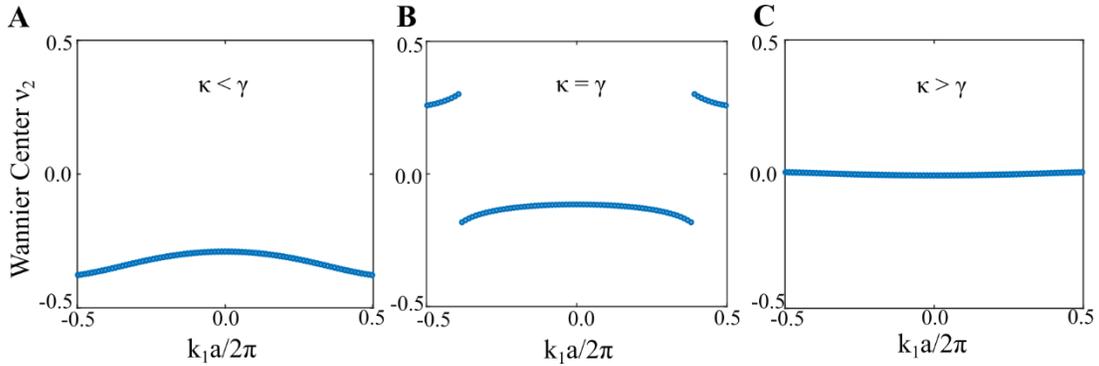

**Figure S1. Wannier bands of Kagome lattice calculated from TBM for the cases of** (A) expanded lattice, (B) unperturbed lattice and (C) shrunken lattice. The ratio of intra-cell and intel-cell hopping terms are (A) $\kappa/\gamma = 0.252$, (B) $\kappa/\gamma = 1$ and (C) $\kappa/\gamma = 3.97$.

## S3 Symmetric constraint on Wannier bands of Kagome lattice

In general, the constraint of the symmetric operator over the Wilson loop satisfies the following relation (2)

$$\boldsymbol{B}_{g,k_j} W_{l_i,k_j} \boldsymbol{B}^\dagger_{g,k_j} = W_{D_g l_i, D_g k_j}, \tag{S7}$$

$l_i$ is the starting point of the path in the Brillouin zone, $\boldsymbol{B}^{n,m}_{g,\boldsymbol{k}} = \langle u^n_{D_g \boldsymbol{k}} | g_{\boldsymbol{k}} | u^m_{\boldsymbol{k}} \rangle$ is the unitary sewing matrix in which the unitary operator $g_{\boldsymbol{k}}$ transforms the Hamiltonian following the formula

$$g_{\boldsymbol{k}} h_{\boldsymbol{k}} g^\dagger_{\boldsymbol{k}} = h_{D_g \boldsymbol{k}}, \tag{S8}$$

and symmetric operator $D_g$ sends $\boldsymbol{k}$ to $D_g \boldsymbol{k}$. For the Kagome lattice, there are 3 reciprocal vectors defined in the Brillouin zone that preserve the $C_3$ symmetry, namely, $\boldsymbol{b}_1 = \frac{4\pi}{\sqrt{3}a_0}(0,1)$, $\boldsymbol{b}_2 = \frac{4\pi}{\sqrt{3}a_0}\left(-\frac{\sqrt{3}}{2}, -\frac{1}{2}\right)$, $\boldsymbol{b}_3 = \frac{4\pi}{\sqrt{3}a_0}\left(\frac{\sqrt{3}}{2}, -\frac{1}{2}\right)$. Therefore, the path $l_i$ can trace along either one of these vectors, and $k_j$ can be the other two vectors.

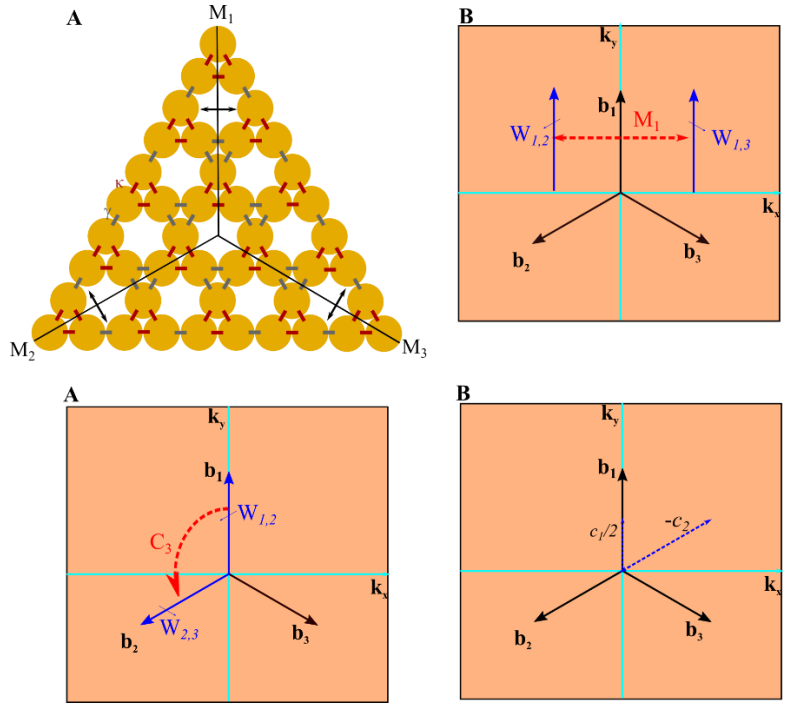

**Figure S2. Schematics of the model and relation between Wilson loop and operators. (A)** Two dimensional Kagome lattice with $C_3$ symmetry and reflection symmetry along $M_1, M_2$ and $M_3$. **(B)** Relation between Wilson loop and reflection symmetry $M_1$. **(C)** Relation between Wilson loop and rotational symmetry $C_3$. **(D)** Relation between Wilson loops traced along path $-c_2$ and $c_1$.

a. Constraint of reflection symmetry $M_i$ over $v_i(k_j)$

There are 3 different reflection symmetric lines for the shrunken or expanded Kagome lattices. Specifically,

$$M_1: (x, y) \to (-x, y),$$

$$M_2: (x, y) \to \left(\frac{1}{2}x + \frac{\sqrt{3}}{2}y, \frac{\sqrt{3}}{2}x - \frac{y}{2}\right),$$

$$M_3: (x, y) \to \left(\frac{1}{2}x - \frac{\sqrt{3}}{2}y, -\frac{\sqrt{3}}{2}x - \frac{y}{2}\right), \tag{S9.1}$$

and for reciprocal vectors

$$M_1: (b_1, b_2, b_3) \to (b_1, b_3, b_2),$$

$$M_2: (b_1, b_2, b_3) \to (b_3, b_2, b_1),$$

$$M_3: (b_1, b_2, b_3) \to (b_2, b_1, b_3). \tag{S9.2}$$

Subsequently, as illustrated in the Fig. S2B, the relation between the Wilson loops under the symmetric operator is

$$B_{M_i,k_j} W_{c_i,k_j} B^\dagger_{M_i,k_j} = W_{c_i,k_t}, \quad (i \neq j \neq t). \tag{S10}$$

The eigenvalues of $W_{c_i,k}$ then satisfy the following

$$M_i: e^{i2\pi v_i(k_j)} = e^{i2\pi v_i(k_t)}. \tag{S11}$$

Thus the Wannier bands meet

$$M_i: v_i(k_j) = v_i(k_t) \mod 1. \tag{S12}$$

Therefore, under the reflection operator $M_i$, bulk polarizations have the relation

$$M_i: p_{i,l_j} = p_{i,l_t}. \tag{S13}$$

Which indicates that the value of polarization doesn't depend on the path of the integration that the Wilson loops go around.

b. Constraint of reflection symmetry $M_i$ over $v_j(k_t)$

Using the relation (S7) and (S9), the Wilson loop has the constraint

$$B_{M_i,k_t} W_{c_j,k_t} B^\dagger_{M_i,k_t} = W_{c_t,k_j}, \quad (i \neq j \neq t). \tag{S14}$$

Thus the Wannier bands meet

$$M_i: v_j(k_t) = v_t(k_j) \mod 1. \tag{S15}$$

Accordingly, the polarization of the lowest energy band is

$$M_i: p_j = p_t. \tag{S16}$$

Which prove that the polarizations of all three directions are the same, which is consistent with our numerical calculation of the bulk polarization in Fig. S1.

c. Constraint of reflection symmetry $M_i$ over $v_j(k_i)$

$$\boldsymbol{B}_{M_i,k_i} W_{c_j,k_i} \boldsymbol{B}^\dagger_{M_i,k_i} = W_{c_t,k_i}, \ (i \neq j \neq t). \tag{S17}$$

This constraint is equivalent to the combination of the previous 2 constraints.

d. Constraint of $C_3$ symmetry over $v_i(k_j)$

Assume the $C_3$ symmetry operator follows the direction $(i \to j \to t)$, therefore, Wilson loop satisfies the relation

$$\boldsymbol{B}_{C_3,k_j} W_{c_i,k_j} \boldsymbol{B}^\dagger_{C_3,k_j} = W_{c_j,k_t}, \tag{S18}$$

Thus the Wannier bands meet

$$C_3: v_i(k_j) = v_j(k_t) \bmod 1. \tag{S19}$$

And due to the equivalence of the three reciprocal lattices, the polarizations of the bands for the three choices are equivalent

$$C_3: p_i = p_j = p_t. \tag{S20}$$

e. Constraint of time reversal symmetry $T$ over $v_i(k_j)$.

The time reversal operator flips the sign of momentum vector due to antilinear property of the operator,

$$T: v_i(k_j) = v_i(-k_j). \tag{S21}$$

Therefore, on the polarization

$$T: p_i(k_j) = p_i(-k_j). \tag{S22}$$

If the direction of path integration in momentum space is reversed, $W_{c_i,k_j} \to W_{-c_i,k_j}$, the Wannier center changes correspondingly

$$v_i(k_j) = -v_{-i}(k_j). \tag{S23}$$

To obtain the bulk polarization from the symmetric constraint, let's first compare $W_{-c_i,k_j}$ and $W_{c_t,k_j}$. If the path $-c_i$ is projected along $c_t$, as seen in Fig. S2B, the Wilson loop will only reach half of $c_t$ while it reaches a full loop of $-c_i$. Combined with Eqs. (S20, S23), we have

$$p_{-i} = \frac{p_t}{2} = -p_i. \tag{S24}$$

Consequently, $3p_i = -1, 0$, or equivalently,

$$p_i = -\frac{1}{3}, 0. \tag{S25}$$

Under the constraint of space symmetry and TR symmetry, the polarization of bands for the Kagome lattice is determined without resorting to the detailed Hamiltonian of the lattice structures.

## S4 The point group form of bulk polarization for the lattice with $C_3$ symmetry

From Eq. (S7), the bulk polarization in the thermodynamic limit ($N_k \to \infty$) can be written as (4)

$$p_q = \frac{1}{2\pi} \int_0^1 dk_1 \int_0^1 dk_2 \, \text{Tr}[A_q(\boldsymbol{k})], \tag{S26}$$

where $A_q(\boldsymbol{k}) = \langle \psi | i\partial_q | \psi \rangle$ is the Berry connection, and integration is performed over the normalized BZ. For the $C_3$ invariant insulator with vanishing Chern number, the bulk polarization can be formulated based on Eq. (S26) into (5)

$$e^{-i2\pi(p_q)} = \prod_{n \in occ} \frac{\theta_n(K)}{\theta_n(K')} = \prod_{n \in occ} \frac{\theta_n(K)}{\theta_n(\Gamma)} \frac{\theta_n(\Gamma)}{\theta_n(K')}, \tag{S27}$$

where $\theta_n(\boldsymbol{k}) = \langle u_n(\boldsymbol{k}) | R_3 | u_n(\boldsymbol{k}) \rangle$ is the expectation value of $C_3$ operator under the eigenvectors $u_n(\boldsymbol{k})$ of Hamiltonian, $n$ is the index of the occupied bands, $K(K')$ and $\Gamma$ are the $C_3$ symmetric invariant points, at which $\theta_n$ is also the eigenvalue of $C_3$ operator, namely, $\exp[(\pm i2\pi)/3]$, or 1. If the time reversal symmetry is preserved, $\theta_n(\Gamma) = \theta_n(\hat{T}\Gamma) = 1$, and $\theta_n(K) = \theta_n(\hat{T}K') = \theta_n(K')^*$. Therefore, the bulk polarization under the constraint of time reversal symmetry is

$$e^{-i2\pi(p_q)} = \prod_{n \in occ} \left( \frac{\theta_n(K)}{\theta_n(\Gamma)} \right)^2, \tag{S28}$$

which is the equivalent to the Eq. (1) in the main text.

## S5 Polarization difference between shrunken and expanded lattices

**Condition**: If the unit cell of the shrunken lattice can be transformed into the corresponding unit cell of the expanded lattice by a series of unitary transformations $\hat{U}_s$ (for example, translational and rotational transformations) thereby swapping the amplitudes of $\kappa$ and $\gamma$;

**Proposition:** the difference between bulk polarizations of shrunken and expanded lattice $\delta p_i$ is nontrivial.

Assume eigenstates of the expanded (shrunken) unit cell are $|u_s(\boldsymbol{k})\rangle$, eigenstates of the shrunken (expanded) unit cell after transformation are $|u'_s(\boldsymbol{k})\rangle = \hat{U}_s |u_s(\boldsymbol{k})\rangle$. Let's define the polarization difference as

$$\begin{aligned} p'_{s,i} &= \frac{i}{2\pi} \int_0^1 dk_1 \int_0^1 dk_2 \, \text{Tr}\langle u'_s(\boldsymbol{k}) | \nabla_i | u'_s(\boldsymbol{k}) \rangle \\ &= \frac{i}{2\pi} \int_0^1 dk_1 \int_0^1 dk_2 \, \text{Tr}\langle u_s(\boldsymbol{k}) | \nabla_i | u_s(\boldsymbol{k}) \rangle + \frac{i}{2\pi} \int_0^1 dk_1 \int_0^1 dk_2 \, \text{Tr} \nabla_i \hat{U}_s \\ &= p_{s,i} + \delta p_i. \end{aligned} \tag{S29}$$

Condition 1 indicates deformed infinite Kagome lattice contains both expanded and shrunk unit cells, so that we have the freedom to switch the unit cell between expanded and shrunken ones. For example, the expanded unit cell marked by solid rhombus can be translated by the vector $\boldsymbol{r_0} = \frac{a_0}{\sqrt{3}}(-\frac{\sqrt{3}}{2},\frac{1}{2})$ in Fig.1 A, and then rotated by $\pi$ around its center ($SU(1)*SO(2)$) to transform into the shrunken unit cell marked by the dashed rhombus. From equation (S3-4), the eigenvalues of the expanded (shrunken) unit cells in the two cases are the same while their eigenvectors are changed due to the unitary transformation under $SU(1)*SO(2)$. Thus, eigenstates after transformation are $|u'_s(\boldsymbol{k})\rangle = U(\boldsymbol{k}\to-\boldsymbol{k})\,e^{i\boldsymbol{k}\boldsymbol{r_0}}|u_s(\boldsymbol{k})\rangle$, and the polarization difference is

$$\delta p_i = -\frac{N_o r_0}{2\pi}\int_{-1}^{0} dk_1 \int_{-1}^{0} dk_2\, b_i \cos\theta, \tag{S30}$$

where $\theta$ is the angle between $\boldsymbol{r_0}$ and $\boldsymbol{b_i}$. $\boldsymbol{r_0}$ is the translational vector, $N_o$ is the number of the occupied bands, and $\boldsymbol{b_i}$ is the reciprocal vector as shown in Fig. 1A, $\boldsymbol{b_1} = \frac{4\pi}{\sqrt{3}a_0}(\frac{\sqrt{3}}{2},-\frac{1}{2})$, and $\boldsymbol{b_2} = \frac{4\pi}{\sqrt{3}a_0}(0,1)$. If only lowest band of Kagome lattice is considered, $\delta\boldsymbol{p} = (\frac{1}{3},\frac{1}{3})$.

We regard the second integral term in (S29) as the gauge potential $A_{const}$, and it preserves the gauge invariance by $\nabla \times A'_{s,i}(\boldsymbol{k}) = \nabla \times (A_{s,i}(\boldsymbol{k}) + A_{const})$. Consequently, there is no physical change between the choices of expanded or shrunken unit cell of infinite lattice. $\delta\boldsymbol{p}$ will play a role if there are boundaries in the Kagome lattice, which break the $C_3$ symmetry of the lattice and makes physical changes between the expanded or shrunk unit cell at the boundaries. As we show in the main text, the physical change is accompanied by the presence of edge states at the cut. Note any lattice with point group symmetry $C_n$ could have a nontrivial difference of wave polarization as long as **condition 1** is met, which is also validated by the work from Dr. Hughes's group (*5*).

### S6 Frequency responses of expanded and shrunken trimers

At the preliminary stage of the fabrication, we fabricated the single expanded and shrunken trimers in Fig. 1 (B-C) without connectors in the in-plane direction, thus the hard wall boundary was applied at the inter-cell couplings and only intra-cell coupling term was expected to play a role in the spectra. We tested the frequency responses of the on-site and off-site probing when the source was placed at the bottom port and the detectors was over the top port of the same cylinder and of the neighbor cylinder, respectively. When not in use, the ports are blocked with printed caps. The frequency responses of both on-site (red colored) and off-site (blue colored) probing are shown in Fig. S3, and the results from experimental measurement and theoretical prediction match very well. Two resonance peaks appear in both spectra (Fig. S3 (A-B)), among which the resonance mode with lower frequency has the in phase oscillating behavior, and the other mode with higher frequency represents the dipolar mode. We observed that the distance of the resonance frequencies for the single expanded trimer is shorter than that for single shrunken trimer, which demonstrates the intra-cell coupling of the expanded trimer is less than that of the shrunken trimer. Based on these initial results, we can use the design of single trimer with outer connectors and extend them into the array of topological nontrivial and trivial lattices.

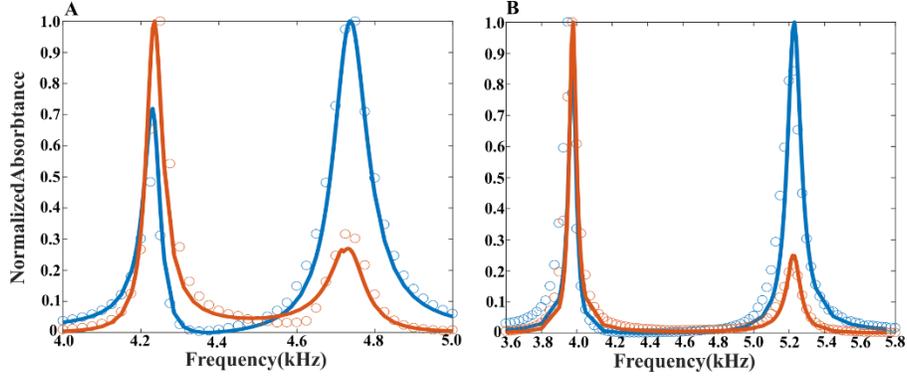

**Figure S3. Normalized absorbance with frequency swept over the range of first-order standing wave of** (A) single expanded trimer and (B) single shrunken trimer terminated by hard wall in the x-y direction, as shown in Fig. 1B-C. Circular dots are the results from acoustic measurement, and solid lines from first principle simulation. Red and blue colors represent the results of on-site and off-site probing, respectively.

## S7 Phase $\phi_{C_3}$ of topological case measured in the expanded and shrunken lattices

Besides carrying out the phase measurement of topological nontrivial and trivial case in the expanded and shrunken lattices, respectively, we also measured the phase differences in the different regions of the expanded and shrunken lattices. As shown in photograph of Fig. S4A, the source was put inside the lattice (yellow spot), and the detectors were placed in the blue dashed region of the expanded lattice, or in the red dashed region of the shrunken lattice, in which cases the same topological phases are obtained as predicted from the polarization difference theorem in section **S6**. The blue and red regions can be transformed into each other by a series of transformation, as described in section **S6**. The corresponding curves of phase $\phi_{C_3}$ extracted from the phase measurement, detailed in the main text, are presented in Fig. S4B, and are in a good agreement with each other. Bulk polarizations are read as $-\frac{1}{3}$ in both cases, combined with the results in Fig. 1I, they reveal that both the topological phase and trivial phase can be measured in the same lattice, thus, confirming our theoretical calculation from polarization difference. Meanwhile, these results also validate our structure can emulate the environment of the infinite lattice for the local detected mode because of the large loss.

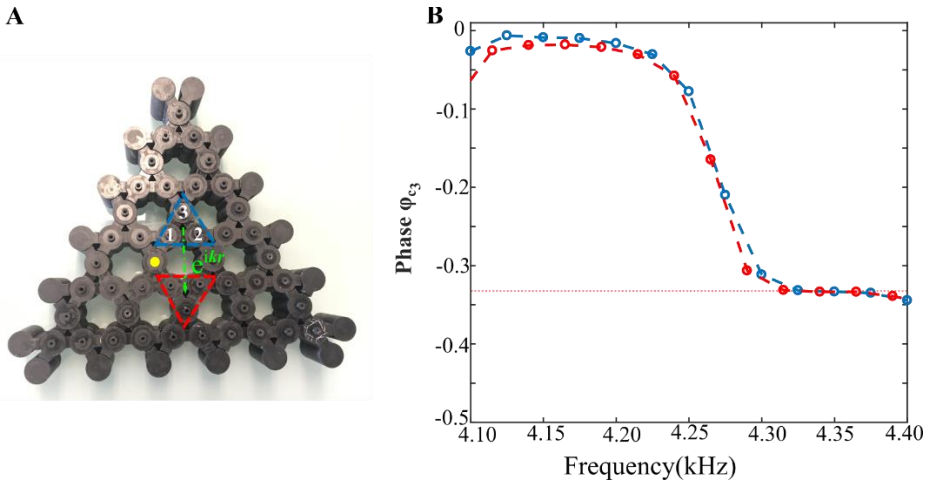

**Figure S4. Experimental demonstration of the equivalence for measuring phase $\phi_{C_3}$ either in the expanded or shrunken lattice.** (**A**) Photograph of the printed acoustic design same as in Fig. 1(**H**), source was placed at yellow spot, and the detectors inside the triangular region of the expanded lattice outlined by blue lines or inside the triangular region of the shrunken lattice outlined by red lines. (**B**) $\phi_{C_3}$ of the lowest bulk band for topological case, blue dashed curve was obtained from the expanded lattice, and red dashed curve obtained from the shrunken lattice. Horizontal dotted line indicates the position $\phi_{C_3} = -1/3$.

## S8 Analytical study of topological edge states in a Kagome strip

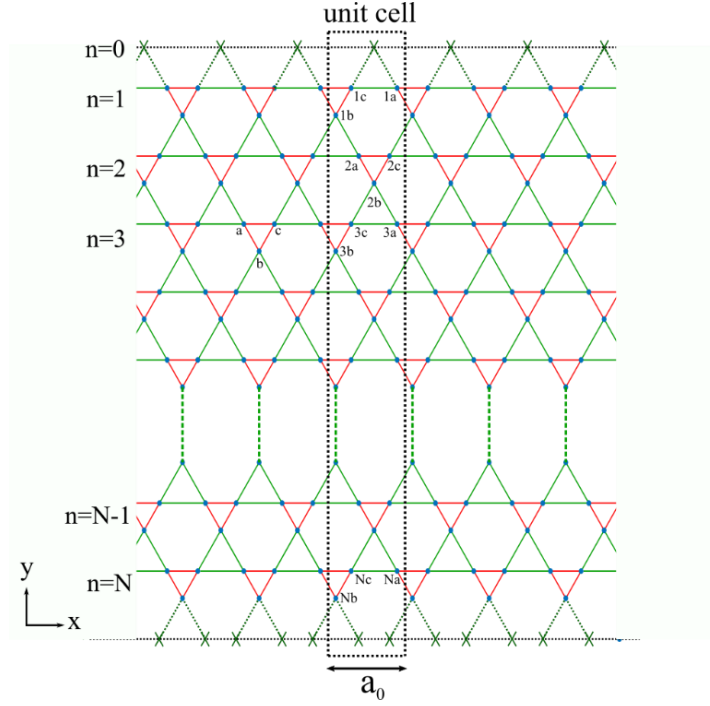

**Figure S5. Schematics of Kagome nanoribbon** terminated along the y-direction, and periodic in the x-direction, as indicated by the rectangular dash lines.

In this section, we consider a nanoribbon periodic in the x direction, and terminated in the y direction, as shown in Figure S5. The nanoribbon consists of N cells in the y direction. Using the TBM we obtain the equations of motion for the nanoribbon:

$$E\psi_{n,a} = -\gamma e^{-ik_x/4}\psi_{n-1,b} - \kappa e^{ik_x/4}\psi_{n,b} - \left(\kappa e^{ik_x/2} + \gamma e^{-ik_x/2}\right)\psi_{n,c},$$

$$E\psi_{n,b} = -\gamma e^{ik_x/4}\psi_{n+1,a} - \kappa e^{-ik_x/4}\psi_{n,a} - \gamma e^{-ik_x/4}\psi_{n+1,c} - \kappa e^{ik_x/4}\psi_{n,c},$$

$$E\psi_{n,c} = -\left(\kappa e^{-ik_x/2} + \gamma e^{ik_x/2}\right)\psi_{n,a} - \gamma e^{ik_x/4}\psi_{n-1,b} - \kappa e^{-ik_x/4}\psi_{n,b}. \quad (S31)$$

where $\psi_{n,q}$ is the probability amplitude of an electron located at site $q = a, b, c$ of the nth unit cell.

With the boundary conditions

$$\psi_{N+1,a} = \psi_{N+1,b} = \psi_{0,c} = 0, \quad (S32)$$

We use a generic solution for the probability amplitudes to be (6):

$$\psi_{n,a} = Ae^{ipn} + Be^{-ipn},$$

$$\psi_{n,b} = Ce^{ipn} + De^{-ipn},$$

$$\psi_{n,c} = Fe^{ipn} + Ge^{-ipn}, \tag{S33}$$

with $A, B, C, D, E, F, G$ as coefficients of the plane waves. The configuration of the cut maintains perpendicularity between $k$ and $p$ wavenumbers. After applying the boundary conditions:

$$\psi_{n,a} = A(e^{ipn} - z^2 e^{-ipn}),$$

$$\psi_{n,b} = C(e^{ipn} - e^{-ipn}),$$

$$\psi_{n,c} = F(e^{ipn} - z^2 e^{-ipn}), \tag{S34}$$

where

$$z^2 = e^{ip(N+1)}. \tag{S35}$$

We substitute these functions into the equations of motion and obtain the relation

$$\mathcal{M}\begin{pmatrix} A \\ C \\ F \end{pmatrix} = 0, \tag{S36}$$

where

$$\mathcal{M}_{\mu,1} = \begin{pmatrix} -E(e^{ipn} - z^2 e^{-ipn}) \\ -\gamma e^{ik_x/4}(e^{ip(n+1)} - z^2 e^{-ip(n+1)}) - \kappa e^{-ik_x/4}(e^{ipn} - z^2 e^{-ipn}) \\ -(\kappa e^{-ik_x/2} + \gamma e^{ik_x/2})(e^{ipn} - z^2 e^{-ipn}) \end{pmatrix},$$

$$\mathcal{M}_{\mu,2} = \begin{pmatrix} -\gamma e^{-ik_x/4}(e^{ip(n-1)} - e^{-ip(n-1)}) - \kappa e^{ik_x/4}(e^{ipn} - e^{-ipn}) \\ -E(e^{ipn} - e^{-ipn}) \\ -\gamma e^{ik_x/4}(e^{ip(n-1)} - e^{-ip(n-1)}) - \kappa e^{-ik_x/4}(e^{ipn} - e^{-ipn}) \end{pmatrix},$$

$$\mathcal{M}_{\mu,3} = \begin{pmatrix} -(\kappa e^{ik_x/2} + \gamma e^{-ik_x/2})(e^{ipn} - z^2 e^{-ipn}) \\ -\gamma e^{-ik_x/4}(e^{ip(n+1)} - z^2 e^{-ip(n+1)}) - \kappa e^{ik_x/4}(e^{ipn} - z^2 e^{-ipn}) \\ -E(e^{ipn} - z^2 e^{-ipn}) \end{pmatrix}.$$

The non-trivial solution for coefficients $A, C, F$ and arbitrary $n$ requires $\det(\mathcal{M})=0$ with $p \neq 0, \pm\pi$ (these solutions are unphysical as it would imply that there's no group velocity across the nanoribbons). After simplifying, we can show that $\det(\mathcal{M})=0$ has the following form:

$$ve^{3ipn} + we^{-3ipn} + se^{ipn} + te^{-ipn} = 0. \tag{S37}$$

Therefore all coefficients $v, w, s, t$ must equal zero. After dividing out co-factors in the coefficients, we obtain the following energy spectrum from the $e^{\pm 3ipn}$ coefficients:

$$(E + \kappa + \gamma)\left(E^2 + E(\gamma + \kappa) - 2(\gamma^2 - \gamma\kappa + \kappa^2) - 2\gamma\kappa(\cos(k) + 2\cos(\tfrac{k_x}{2})\cos(p))\right) = 0,$$
(S38)

yielding

$$E = -\kappa - \gamma,$$
$$E = \tfrac{1}{2}(-\gamma - \kappa \pm g(k_x, p)), \tag{S39}$$

with

$$g(k_x, p) = \sqrt{9\gamma^2 - 6\gamma\kappa + 9\kappa^2 + 8\gamma\kappa\cos(k_x) + 16\gamma\kappa\cos(\tfrac{k_x}{2})\cos(p)}.$$

The eigenvalues obtained above are the same as those calculated from TBM in Eqs. (S3). Similarly, from the $e^{\pm ipn}$ coefficients (after controlling for cofactors) and substituting in the eigenvalue Eqs. (S39), we obtain the equation relating $p$, $k_x$ and $N$:

$$F(p, N) \equiv \gamma\bigl(3\gamma + \kappa + 2\kappa\cos(k_x) + g(k_x, p)\bigr)\sin(Np)$$
$$+ \kappa\cos(\tfrac{k_x}{2})\bigl(3\gamma + 3\kappa + g(k_x, p)\bigr)\sin((1 + N)p) = 0. \tag{S40}$$

For $\gamma > \kappa$ and arbitrary $k_x \in (0, 2\pi)$, the equation above will always return $N - 1$ solutions for $p$ (Fig. S6A). However, for $\gamma < \kappa$ and given $k_x$, the above equation will return either $N$ or $N - 1$ dependent on some critical value of $k_x$, $k_c$. When the slope of $F(p, N)$ approaches to 0, one of the roots will vanish, this is the condition for the edge states to exist and can be found numerically by solving the following equation:

$$\tfrac{\partial}{\partial p}F(p, N)\big|_{p=\pi} = 0. \tag{S41}$$

As we can see in the example of $N = 5$, $k_c$ is close to $\pi$ and the edge states bifurcate from the bulk states in the region near $k_x = \pi$ for the shrunken lattice, as shown in our previous work (7). Edge states in this case are present due to the $C_2$ symmetry instead of $C_3$ symmetry, therefore, the shrunken lattice is still regarded as trivial insulator with respect to the bulk polarization arising from $C_3$ symmetry.

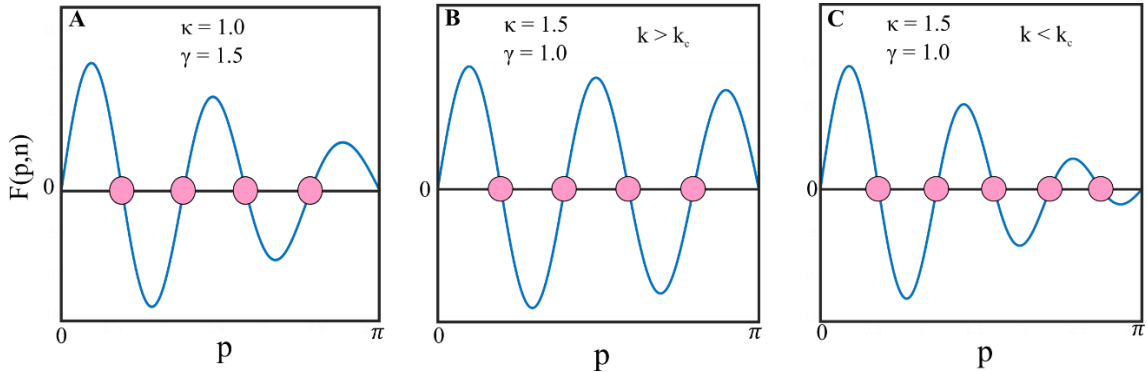

**Figure S6. Determine the existence of edge states by counting the number of roots of $F(p,N)$.** (A) Curve of $F(p,N)$ for expanded lattice with $\gamma > \kappa$ and $N=5$. There are always N-1 solutions and the missing state corresponds to the edge state. (B) Curve of $F(p,N)$ for the shrunken lattice with $\gamma < \kappa$ and $k < k_c$, where $k_c \cong 0.86\pi$. All states are accounted for in the bulk and there is no edge state. (C) Curve of $F(p,N)$ for the shrunken lattice with $\gamma < \kappa$ and $k > k_c$. The Nth bulk state disappears and an edge state is present instead.